\documentclass[12pt]{article} 
\usepackage{graphicx}
\newcommand{\be}{\begin{equation}}
\newcommand{\ee}{\end{equation}}
\newcommand{\bear}{\begin{eqnarray}}
\newcommand{\eear}{\end{eqnarray}}
\newcommand{\ba}{\begin{array}}
\newcommand{\ea}{\end{array}}

\begin{document}

\begin{titlepage}
\vfill
\begin{flushright}
{\normalsize IC/2009/001 }\\
\end{flushright}

\vfill
\begin{center}
{\Large\bf Fate of $Z_N$ walls in hot holographic QCD}

\vskip 0.3in

{ Ho-Ung Yee\footnote{\tt hyee@ictp.it}}

\vskip 0.15in

 {\it ICTP, High Energy, Cosmology and Astroparticle Physics,} \\
{\it Strada Costiera 11, 34014, Trieste, Italy}
\\[0.3in]

{\normalsize  2008}

\end{center}

\vfill

\begin{abstract}
\normalsize\noindent
We first study $Z_N$ walls/interfaces in a deconfined phase of Witten's $D4$-brane background of pure $SU(N)$
Yang-Mills theory, motivated by a recent work in the case of $N=4$ SYM. Similarly to
it, we propose that
for a large wall charge $k\sim N$, it is described by $k$ $D2$-branes blown up into
a $NS5$-brane wrapping $S^3$ inside $S^4$ via Myers effect, and we calculate the tension by
suitable U-duality. We find a precise Casimir scaling for the tension formula.
We then study the fate of $Z_N$-vacua in a presence of fundamental flavors
in quenched approximation via gauge/gravity correspondence. In the case of $D3/D7$ system
where one can vary the mass $m_q$ of flavors,
we show that there is a phase transition at $T_c\sim m_q$, below which the $Z_N$-vacua survive
while they are lifted above the critical temperature. We analytically calculate
the energy lift of $k$'th vacua in the massless case, both in the $D3/D7$ system
and in the Sakai-Sugimoto model.
\end{abstract}

\vfill

\end{titlepage}
\setcounter{footnote}{0}

\baselineskip 18pt \pagebreak
\renewcommand{\thepage}{\arabic{page}}
\pagebreak

\section{Introduction }

Gauge/gravity correspondence has often provided us easy tools for addressing
difficult problems of strongly interacting gauge theories.
It is based on the observation that large N, large t'Hooft coupling limit
of certain gauge theories have alternative weakly-coupled effective descriptions
in terms of gravity or string theory \cite{Maldacena:1997re}.
One important aspect of these dual theories is that the spacetime they are living
is one-dimension higher than the original spacetime of gauge theories, and
the additional dimension, often referred to as holographic coordinate,
plays a physical role of energy scale of a given process.
This mapping of space-energy is practically achieved by the presence of
a warping factor in the metric that depends on the additional dimension, which
provides a geometrical potential along the holographic dimension.

Out of vast amount of previous works
that have been done over a decade to test and study this idea,
one interesting application has been to describing the deconfined phase
of gauge theories in a sufficiently high temperature, whose
dual picture is a black-hole spacetime with its horizon located at some point
in the holographic coordinate \cite{Witten:1998qj,Witten:1998zw}. Due to the geometric barrier along
holographic dimension provided by
the warping factor in the metric, this black-hole spacetime is stable
against its Hawking radiation, resulting in an equilibrium state called Hartle-Hawking state.
Performing relatively easy analyses in this black-hole background
has given us many valuable, typically non-trivial, results about physics
of high temperature, deconfined gauge theories in strong coupling \cite{Policastro:2001yc}.

In this work, we would like to add one more piece to the
mounted pile of results about high temperature phase
of gauge theories obtained using gauge/gravity correspondence.
Our motivation is two-fold.
We first study $Z_N$ walls in the Euclidean picture of finite temperature
deconfined phase of pure $SU(N)$ Yang-Mills theory using the Witten's dual background of
$D4$-branes \cite{Witten:1998zw}. We are interested in the $k$-wall with ${k\over N}\sim {\cal O}(1)$,
which is largely motivated by a recent work of Armoni, Kumar and Ridgway for the case of
$N=4$ SYM via the $AdS_5\times S^5$ background \cite{Armoni:2008yp}.
Similar to their result, we will show that the $k$-wall with ${k\over N}\sim {\cal O}(1)$
is $k$ (Euclidean) $D2$-branes blown up to a single $NS5$-brane wrapping $S^3$
inside the $S^4$ in the background via Myers effect.
We calculate its tension after performing suitable U-dualities of Type II SUGRA
to make the problem more tractable.
Our result of $k$-wall tension indicates certain limitations of finite temperature
black-hole geometry of
Witten's background that has been used in the previous literature.

Our second motivation is to study the fate of $Z_N$ discrete vacua, and hence
$Z_N$ walls connecting them,
when we include fundamental flavor quarks in the gauge theory.
One naturally expects that since $Z_N$ is no longer a symmetry of the theory
in the presence of fundamental matters, these $Z_N$ vacua are generally lifted
except the one true vacuum.
However, there appears an issue when the fundamental matters are much more massive than
the temperature, because the physics with energy scale below the mass of
fundamental matters decouples from the fundamental matters, and the effective
low energy theory would still have a center $Z_N$ symmetry.
One logical possibility is that when ${m_q\over T}\gg 1$, the $Z_N$ vacua and domain walls
persist in the theory, while they disappear at ${m_q\over T}\ll 1$, and these
two phases are separated by a phase transition at ${m_q\over T_c}\sim {\cal O}(1)$.
We will give an evidence to this scenario in the framework of
gauge/gravity correspondence using $D3/D7$-brane system corresponding to
$N=2$ fundamental hypermultiplets in $N=4$ SYM \cite{Karch:2002sh}.

In the case of massless fundamental flavors, $Z_N$ vacua are lifted unambiguously.
Using gauge/gravity correspondence, we
calculate analytically this energy lift of $k$-vacua in quenched approximation,
in both $D3/D7$ system and in the Sakai-Sugimoto model \cite{Sakai:2004cn}.
As a by-product, we also find an infinite tower of stable states above
each lifted $k$-vacua, similar to the tower of stable states above $\theta$-vacuum
that was found in Ref.\cite{Witten:1998uka}.

{\it Note added in revision: }

The author was informed of an issue of correct interpretation of center $Z_N$ symmetry and
the corresponding $Z_N$ walls in Euclidean thermal $SU(N)$ gauge theory\footnote{We thank Andrei Smilga
for bringing this issue to our attention.}.
If one takes $SU(N)/{Z_N}$ as the gauge group rather than $SU(N)$, the center $Z_N$ symmetry is absent and
the $Z_N$ walls are not the Minkowski domain walls, but instead some Euclidean instanton-like objects \cite{Smilga:1993vb}.
In the presence of fundamental flavors, the gauge group is indeed $SU(N)$ and the center $Z_N$ symmetry
can be at least an approximate symmetry of the theory. Our second subject of the paper should be
viewed in this way. Alternatively, one can think of thermal $S^1$ as a spatial compactification, and
consider 3-dimensional Minkowski theory after Wick rotation along one non-compact spatial direction.
In that context, one can talk about Minkowski domain walls in the 3 dimensional theory.

\section{Review of center $Z_N$ in gauge/gravity correspondence}

This section is devoted to a brief review on the center $Z_N$ symmetry of
the Euclidean description of finite temperature $SU(N)$ gauge theory with only
adjoint matters, and its spontaneous breaking in the deconfined phase via
non-vanishing Polyakov line along the thermal circle \cite{Yaffe:1982qf,Svetitsky:1982gs}.
We will also summarize relevant results in Refs.\cite{Armoni:2008yp,Aharony:1998qu}
in the context of gauge/gravity
correspondence
to lay ground for our analyses
in subsequent sections.
Readers familiar to these subjects can jump into the next section.

\subsection{Center $Z_N$ and its spontaneous breaking in hot $SU(N)$ gauge theory}

An Euclidean finite temperature gauge theory is defined in the space $S^1\times R^3$,
where the thermal circle, or Euclidean time, has a period $\beta={1\over T}$.
As a $SU(N)$ gauge theory, field configurations that are connected by $SU(N)$ gauge
transformations must be treated as identical to each other, and one should mod out
the space of fields by gauge transformations.
Having $S^1$ means that the local gauge function $U(x)\in SU(N)$ that we mod out
should be periodic along $S^1$ to preserve usual periodicity of matter fields
that are charged under $SU(N)$.
If the theory contains only adjoint matters, one might think of an extended notion of
gauge transformation whose gauge function $U(x) \in SU(N)$ is periodic
only up to an element of the center $Z_N$ inside $SU(N)$.
This is because for adjoint matters $\Phi$, as $Z_N$ freely commutes with adjoint $\Phi$,
the gauge transformation
\be
\Phi(x) \to U(x)^\dagger \Phi(x) U(x)\quad,
\ee
doesn't change the periodicity of $\Phi$, and seems to be allowed without any problem.
Note however that under the extended gauge transformation, the Wilson line along
the thermal $S^1$, called Polyakov line,
\be
W(S^1)={1\over N}{\rm Tr}P\exp\left(i\int_{0}^\beta A_\tau d\tau\right)\quad,
\ee
transforms exactly by the center $Z_N$ element $W(S^1)\to e^{2\pi i k\over N} W(S^1)$.
Therefore, if we accept $W(S^1)$ as one of physical "gauge invariant" observables of the theory,
the extended gauge transformations are in fact not allowed as gauge transformations
that we mod out. Rather they should be thought of as symmetries of the theory as the action
is clearly invariant under the transformations.
The extended gauge transformations are constant $Z_N$ times the usual periodic gauge transformations
that we mod out, so that the resulting symmetry is a global $Z_N$ symmetry.

One can immediately identify $W(S^1)$ as an order parameter of the spontaneous breaking
of this center $Z_N$ symmetry. In fact, it can be argued to get a non-vanishing
expectation value in the deconfined phase of high temperature, while it should vanish
in the confined phase. The rough picture is that $W(S^1)$ represents
a world-line of an external fundamental quarks sitting at a point in $R^3$
in Euclidean finite temperature description. In the confined phase, its presence
would cost too much free energy due to confinement
so that the partition function with $W(S^1)$ would vanish,
while one generally expects a finite non-vanishing result in the deconfined phase.
Therefore, $Z_N$ is spontaneously broken in the deconfined phase, and
there exist $N$-number of vacua parameterized by the $Z_N$ phase of
the order parameter $W(S^1)$. One naturally thinks of domain walls
connecting these vacua. By $k$-wall, we will refer to a domain wall
connecting $i$'th vacuum and $i+k$'th vacuum, and $k$ is identified with
$k+N$ by definition.

There is a beautiful connection between a $Z_N$ domain wall and
a large spatial t'Hooft line \cite{KorthalsAltes:1999xb}.
Consider a large spatial t'Hooft line along a curve $C$ which bounds a large surface $S$
spanning two space directions in $R^3$, say $x^1$ and $x^2$.
By t'Hooft line, we mean inserting
a magnetic monopole whose world-line is the curve $C$. In the presence
of a magnetic monopole, the Bianchi identity for the electric gauge potential $A$
is violated by
\be
dF=\delta^{(3)}_C\quad,
\ee
where $\delta^{(3)}_C$ is the Poincare dual 3-form to $C$.
Being familiar to the case of Dirac monopole, we can think of $S$ as a Dirac sheet, a world-sheet
spanned by a Dirac string.
As $\partial S=C$,
we have
\be
d\delta^{(2)}_S=\delta^{(3)}_C\quad,
\ee
and we see that
\be
F=dA=\delta^{(2)}_S\quad.\label{fda}
\ee
Then, consider a cylinder $D$ made of $S^1$ times a finite interval in $R^3$ which connects
two points $P,Q$ from the two opposite sides of the surface $S$, so that this interval crosses
$S$ at a single point in $R^3$. It is clear that this cylinder has an intersection number 1
with the surface $S$ in our spacetime $S^1\times R^3$, or equivalently
\be
\int_D \delta^{(2)}_S=\#(D,S)=1\quad.
\ee
Using (\ref{fda}), the left-hand side in the above is
\be
\int_D F=\int _{\partial D} A = \int_{S^1 at P} A -\int_{S^1 at Q} A \quad,
\ee
which tells us that the phase of Polyakov line jumps across the surface $S$ of Dirac sheet,
and we can think of $S$ as a domain wall separating regions of different Polyakov lines.
Although the above discussion is given in terms of an Abelian gauge theory,
the logic is essentially identical in the case of non-Abelian $SU(N)$ gauge theory,
and the conclusion is that the $k$-wall can be thought of
as the minimal surface $S$ bounded by
a large spatial
t'Hooft line of $k$-th anti-symmetric representation of the magnetic group.

\subsection{$Z_N$ vacua and $Z_N$ walls in gauge/gravity correspondence}

As one typically considers the large N limit in discussing
gauge/gravity correspondence, it may at first
seem unlikely to see discrete $Z_N$ symmetry in the gravity dual description.
However, believing AdS/CFT correspondence beyond the leading large N limit,
one should be able to access sub-leading $1\over N$ effects by carefully
taking into account quantum effects in the gravity side, which is not easy in general.
In the case of $Z_N$, Aharony and Witten successfully identified the relevant
quantum effect in $AdS_5\times S^5$ background,
that is responsible for the discrete $Z_N$ symmetry
appearing in the gravity dual description of the $N=4$ SYM \cite{Aharony:1998qu}.

For our purposes, we will recapitulate only the case of Poincare patch version
of $AdS_5\times S^5$ corresponding to the gauge theory defined on $S^1\times R^3$,
while we refer the readers to Ref.\cite{Aharony:1998qu}
for the discussion of $N=4$ SYM theory on $S^1\times S^3$.
This is because only in the former case, we can have a spontaneous breaking of global
symmetry such as $Z_N$ that we are interested in.
A finite temperature phase of $N=4$ SYM in Euclidean description has a unique
dual geometry given by the Euclideanized black-hole in the Poincare patch;
\be
ds_E^2={r^2\over L^2}\left[\left(1-{\pi^4 T^4 L^8\over r^4}\right) dt_E^2 +dx_i^2\right]
+{L^2\over r^2}{1\over \left(1-{\pi^4 T^4 L^8\over r^4}\right)}dr^2 +L^2 d\Omega_5^2\quad,\label{n=4decon}
\ee
where the thermal circle combined with the holographic radial coordinate $r$
makes a two-dimensional cigar shape $D$ that closes off at the location of the horizon
$r_H=\pi T L^2$ with $L^4=4\pi g_s N l_s^4$.
The theory is always in the deconfined phase due to conformal symmetry.
As is well-known, the expectation value of a Wilson line is calculated
by the semi-classical string world-sheet which has a boundary at UV $r\to\infty$
along the curve of the Wilson line \cite{Rey:1998ik,Maldacena:1998im}.
One easily identifies such string world-sheet in the case of our
Polyakov temporal Wilson line $W(S^1)$; it spans precisely the cigar $D$ of the thermal circle and $r$.
Because the world-sheet closes off at a finite point $r=r_H$, it would give us a finite
value of the string action,
and hence non-vanishing expectation value of $W(S^1)$ after suitable holographic renormalization \cite{de Haro:2000xn},
which is in accord with the fact that $W(S^1)$ is non-vanishing in a deconfined phase.
However, there appears a puzzle since it seems one can freely turn on NS 2-form $B$ with $dB=0$
on the cigar without any change in the Type IIB SUGRA action, with arbitrary values of
\be
{1\over 2\pi l_s^2}\int_D B\quad,
\ee
which appears as a phase of our semi-classical string amplitude, and hence of $W(S^1)$.
As this $B$ mode is a normalizable mode, we in fact have to sum over these possible phases
in the partition function, which would make the expectation value of $W(S^1)$ zero \cite{Witten:1998zw}.
To prevent this, there must be a mechanism that lifts the degeneracy among continuous values of
NS $B$ field, and we indeed also need this to have $Z_N$ discrete values of the phase of $W(S^1)$.

Ref.\cite{Aharony:1998qu} showed that the necessary mechanism is from quantization of the RR 2-form $C$
in our background with $N$ D3 flux on $S^5$,
\be
{1\over (2\pi l_s)^4}\int_{S^5} F_5 = N\quad,
\ee
such that the action contains the piece
\be
N\cdot \left({1\over 2\pi l_s^2}\int_D B\right) \cdot{1\over (2\pi l_s)^2}\int_{R^3} dC
\equiv \theta \cdot{1\over (2\pi l_s)^2}\int_{R^3} dC\quad,\label{zNinads}
\ee
from the Type IIB term ${2\pi\over (2\pi l_s)^8}\int_{M_{10}} F_5\wedge B_2\wedge H_3$ with $H=dC$.
With the usual kinetic term for $H=dC$ along $R^3$
obtained after integrating out $S^5\times D$,
\be
{g_s N\over 2^7\pi^5 l_s^4 T^3}\int_{R^3}\,\left(H_{123}\right)^2\equiv
{1\over 2 e^2}\int_{R^3}\,\left(H_{123}\right)^2\quad,
\ee
the above $\theta$ was shown to play
a similar role as the ${\theta\over 2\pi} F_{01}$ term in 2D Abelian
$U(1)$ theory, where the vacuum energy is given by
\be
E_{\theta}={e^2\over 2(2\pi l_s)^4} \cdot {\rm min}_{k\in Z}(2\pi k-\theta)^2\quad,
\ee
with integer $k$ labelling certain allowed quantum states of the system.
This says $\theta=2\pi Z$ are discrete vacua among continuous possibilities, and
looking back (\ref{zNinads}), one readily finds discrete $Z_N$ vacua of
\be
{1\over 2\pi l_s^2}\int_D B={2\pi k \over N}\quad,\quad k\sim k+N\quad,
\ee
in the quantum theory, which at the same time also explains
the $Z_N$ phase of $W(S^1)$ at each vacua.

The $Z_N$ walls in the gravity description can most easily be
identified by considering their connection to large spatial t'Hooft lines that
we discussed above.
In the simplest case of $k=1$ wall, the t'Hooft line corresponds
to a (Euclidean) $D1$ brane world-sheet whose boundary at $r\to\infty$ is along the
t'Hooft line curve. One may think of the $D1$ brane as the Dirac sheet of the t'Hooft line,
and one concludes that $D1$ brane spanning along two spatial directions in $R^3$ is the
$k=1$ domain wall. Note that when the t'Hooft line is large, most of the $D1$
world-sheet will sit at the IR end $r=r_H$ to minimize its tension.
Therefore, the $D1$ brane sitting at $r=r_H$ and a point in $S^5$, while spanning
two directions in $R^3$ is the final configuration we are seeking for the $k=1$ wall in $N=4$ SYM.
In a recent work by Armoni, Kumar and Ridgway \cite{Armoni:2008yp}, they proposed that for $k$-walls
with ${k\over N}\sim{\cal O}(1)$,
the $k$ $D1$ branes are blown up to a $NS5$ brane wrapping $S^4\subset S^5$ via Myers effect,
and the result for the $k$-wall tension in their picture supports this claim, that is,
for low $k$ the tension is simply that of the $k$ $D1$ branes, and $(N-k)$-wall
tension is equal to the $k$-wall tension.

\section{ $k$-walls in the Witten's $D4$-brane background}

We come to one of our objectives in this work; calculating $k$-wall tension
with ${k\over N}\sim {\cal O}(1)$ in the (approximate)
gravity dual of pure $SU(N)$ Yang-Mills theory.
We first discuss briefly how center $Z_N$ appears in the Witten's background,
which is rather closely parallel to the case of $AdS_5\times S^5$.

There are two finite temperature Euclidean geometries
competing with each other near the confinement-deconfinement phase transition; a confined phase at low $T$
where the thermal circle remains finite in all spacetime region, and
a deconfined phase of Euclideanized black-hole where the thermal circle shrinks
to zero at the horizon, making a cigar shape combined with the holographic direction \cite{Aharony:2006da}.

The confined phase geometry is
\bear
ds^2&=&\left(U\over R\right)^{3\over 2}\left(dt_E^2+dx_i^2+f(U)dx_4^2\right)
+\left(R\over U\right)^{3\over 2}\left({dU^2\over f(U)}+U^2 d\Omega_4^2\right)\quad,\nonumber\\
F_4&=&{(2\pi l_s)^3 N\over V_4}\epsilon_4\quad,\quad e^\phi=g_s\left(U\over R\right)^{3\over 4}\quad,
\quad V_4={\rm Vol}(S^4)={8\pi^2\over 3}\quad,\nonumber\\
R^3&=& \pi g_s N l_s^3\quad,\quad f(U)=1-\left(U_{KK}\over U\right)^3\quad,\label{confinedgeo}
\eear
with the thermal circle $t_E$ periodic $t_E\sim t_E+\beta$, and $x_4$ is compactified
with the period
\be
\delta x_4= {4\pi\over 3}\left(R^3\over U_{KK}\right)^{1\over 2}\equiv {2\pi\over M_{KK}}\quad.
\ee
The useful relation between the above parameters and the 4D gauge theory data on $(t_E,x_i)$
is given by \cite{Sakai:2004cn}
\be
R^3={g_{YM}^2 N l_s^2\over 2 M_{KK}}\quad,\quad
U_{KK}={2\over 9}g_{YM}^2 N M_{KK} l_s^2\quad,\quad
g_s={g_{YM}^2\over 2\pi M_{KK} l_s}\quad.
\ee
As usual the Polyakov line expectation value is described by a string world-sheet
wrapping $t_E$ and one more direction in the geometry, and since $t_E$ never shrinks
to zero one easily finds that there is no way this string stops in the geometry
without turning back to the UV boundary $U\to\infty$, which corresponds to another
anti-Polyakov line in the gauge theory. In short, $\langle W(S^1)\rangle=0$ with
a single insertion of $W(S^1)$ in the gauge theory, in accord to the expectation in a
confined phase. The way discrete $Z_N$ symmetry appears in the situation is that
if we insert $W(S^1)$'s by multiple N times, these multiple N number of
string world-sheets can now sit on (Euclidean) $D4$ branes wrapping $S^4\times t_E$,
because on this baryonic $D4$-brane, there is a Chern-Simons coupling
\be
\mu_4\int_{D4} F_4\wedge(2\pi l_s^2)A ={1\over (2\pi l_s)^3}\int_{D4} F_4\wedge A
=N\int dt_E \,A_0\quad,
\ee
which induces $N$ string charges on the world-volume, so that N strings can sit on it \cite{Witten:1998xy}.
Therefore, multiple N times insertion of $W(S^1)$'s gives us a finite expectation value
and one concludes that there is a $Z_N$ symmetry under which $W(S^1)$
has a unit charge. Note that the $Z_N$ symmetry is unbroken in this phase due to $\langle W(S^1)\rangle=0$.

Our present interest concerns more about the deconfined phase at high $T$,
for which the dual geometry is given by
\bear
ds^2&=&\left(U\over R\right)^{3\over 2}\left(f(U)dt_E^2+dx_i^2+{1\over (M_{KK}l_s)^2}dx_4^2\right)
+\left(R\over U\right)^{3\over 2}\left({dU^2\over f(U)}+U^2 d\Omega_4^2\right)\quad,\nonumber\\
F_4&=&{(2\pi l_s)^3 N\over V_4}\epsilon_4\quad,\quad e^\phi=g_s\left(U\over R\right)^{3\over 4}\quad,
\quad V_4={\rm Vol}(S^4)={8\pi^2\over 3}\quad,\nonumber\\
R^3&=& \pi g_s N l_s^3\quad,\quad f(U)=1-\left(U_T\over U\right)^3\quad,\label{decon}
\eear
where the period of the thermal circle, $\beta={1\over T}$, is related to $U_T$ by
\be
\beta={4\pi\over  3}\left(R^3\over U_T\right)^{1\over 2}\quad,
\ee
and we have re-scaled $x_4$ such that its period is $(2\pi l_s)$ for later convenience.
In this background, the string world-sheet for a Polyakov line
can stop at $U=U_T$ where the thermal circle $t_E$ shrinks to zero making a cigar shape $D$ with $U\ge U_T$,
so that the string action becomes finite and one expects a non-vanishing VEV of $W(S^1)$.
The same issue of $NS$ 2-form phase ${1\over 2\pi l_s^2}\int_D B^{NS}$
arises as in the case of $AdS_5\times S^5$, and the resolution is also similar.
In Type IIA, there is a term
\be
{2\pi\over (2\pi l_s)^8}\int_{M^{10}} F_4\wedge F_4\wedge B_2
=N\cdot \left({1\over 2\pi l_s^2}\int_D B^{NS}\right){1\over (2\pi l_s)^3}
\int_{x_i,x_4} F_4 \equiv \theta\cdot
{1\over (2\pi l_s)^3}
\int_{x_i,x_4} F_4 \quad,
\ee
and considering quantizing $F_4^{RR}$ along $(x_i,x_4)$ after integrating over $S^4\times D$
to get the kinetic term
\be
{3 g_s N\over 2^9 \pi^6 l_s^5 T^3} \int dx_i dx_4\, \left(F_{1234}\right)^2\quad,
\ee
we have the integer parameterized vacua of $\theta=2\pi Z$ to minimize the energy of the allowed
quantum states treating $x_1$ as a time,
\be
\Psi_k=\exp\left(i k \mu_2\int_{x_{2,3,4}} C^{RR}_3\right)=
\exp\left({2\pi i k \over (2\pi l_s)^3}\int_{x_{2,3,4}} C^{RR}_3\right)
\quad.
\ee
This provides us the $Z_N$ vacua of
\be
{1\over 2\pi l_s^2}\int_D B^{NS}={2\pi k \over N}\quad,\quad k\sim k+N\quad,
\ee
corresponding to the spontaneous breaking of $Z_N$ by $\langle W(S^1)\rangle \neq 0$.

It is not difficult to identify the string theory object that plays a role of
$Z_N$ domain walls, at least for $k$-walls with sufficiently low $k$'s.
Again, using the large spatial t'Hooft line is most convenient.
In our $D4$ background, the magnetically charged sources compared to the electrical
degrees of freedom of fundamental strings are provided by $x_4$ wrapping $D2$-branes \cite{Brandhuber:1998er}.
This can also be seen by considering T-duality along $x_4$ upon which
N $D4$-branes become $D3$-branes and $D2$ becomes $D1$, the usual magnetic degrees of freedom.
Consequently, a large spatial t'Hooft line will be the boundary of the world-volume
of a $x_4$-wrapped (Euclidean) $D2$-brane at $U\to\infty$, and
most of the $D2$-brane world-volume will sit at the IR end $U=U_T$ to minimize its tension.
We can think of two spatial directions in $R^3$ that $D2$-brane is spanning
as the Dirac sheet corresponding to the t'Hooft line, and hence the $k=1$ domain wall.
Therefore, the $D2$-brane sitting at $U=U_T$ and a point in $S^4$ while
spanning $x_4$ and two directions in $R^3$ is the desired $k=1$ domain wall in the gravity side.
For small number of $k$, one naturally expects a collection of $k$ $D2$-branes
to be the corresponding object for the $k$-wall.
Its tension is calculated straightforwardly by the DBI action of $D2$-branes at $U=U_T$,
\bear
T_{k}=k \mu_2 \int dx_4 \, e^{-\phi}\sqrt{g^*}\,\Bigg|_{U=U_T}=
{2\pi k\over (2\pi l_s)^3}{2\pi\over g_s M_{KK}}\left(U_T\over R\right)^{3\over 2}
= {32\pi^3 \over 27 }kN{T^3\over M_{KK}} \quad.\label{kd2}
\eear
This result is qualitatively different from both the weak-coupling calculations of pure $SU(N)$
Yang-Mills theory \cite{Bhattacharya:1990hk,Bhattacharya:1992qb},
and the recent weak/strong coupling calculations of $N=4$ SYM by Armoni, Kumar and Ridgway \cite{Armoni:2008yp},
where the results have a common form of
\be
T_k =
(const)\cdot  k N {T^2\over \sqrt{g_{YM}^2 N}}\quad,
\ee
for small ${k\over N}\ll 1$.

One can understand the origin of the discrepancy
as follows. The confinement-deconfinement phase transition in the Witten's background
happens at a temperature $T_c$ of order $M_{KK}$ \cite{Aharony:2006da},
and the above deconfinement background
becomes relevant when $T$ is much larger than $M_{KK}$.
However, $M_{KK}$ also serves as a compactification scale of $x_4$ below which
we have a 4D gauge theory, while above which massive KK modes
enter the dynamics and the theory becomes effectively 5-dimensional.
The number of KK modes that would enter at the temperature $T$ is roughly given by $T\over M_{KK}$,
and assuming each new degrees of freedom contributes to the tension of the domain wall,
one can understand ${T^3\over M_{KK}}=T^2\cdot {T\over M_{KK}}$ behavior of our result
for the tension formula. However, missing $\sqrt{g_{YM}^2 N}$ factor remains still puzzling.
One possibility is that the mass of KK modes might become $M_{KK}\over \sqrt{g_{YM}^2 N}$
at strong coupling, and the number of effective degrees of freedom at $T$
might be ${T\over M_{KK}}\sqrt{g_{YM}^2 N}$ instead of $T\over M_{KK}$. Our result
may be considered as pointing out this phenomenon.
The above discussion indicates that the deconfinement geometry (\ref{decon}) that
has been used in the previous literature has some limitation to be used as
a gravity dual background of a 4D gauge theory in its deconfined phase.

We proceed by studying $k$-walls with $k$ being large and comparable to $N$,
motivated by a recent work of Armoni, Kumar and Ridgway in $AdS_5\times S^5$ \cite{Armoni:2008yp}, and
our method in this regard will be similar to theirs.
The naive picture of $k$-wall as a simple collection of $k$ $D2$-branes
should break down when $k\sim N$ because $k$ is $Z_N$-valued and moreover
$k$ is related to $(N-k)$ by charge conjugation and the tension must be invariant under this.
Similar to the claim in Ref.\cite{Armoni:2008yp}, we propose that
$k$ $D2$-branes sitting at a point in $S^4$ blow up into a single IIA $NS5$-brane
wrapping $S^3\subset S^4$ due to the background $F_4$ flux on $S^4$,
and we confirm this picture by computing the $k$-wall tension and check
the necessary properties.

To analyze more easily the Type IIA $NS5$-brane dynamics, with $k$ $D2$-brane charges on its
world-volume along $x_4$
and two spatial directions in $R^3$, say $x_1$ and $x_2$,
we perform U-dualities of Type II theories in the following way.
Note first that in the deconfined phase geometry (\ref{decon}),
$x_4$ size remains finite in all region of the spacetime, contrary to the confined phase geometry
where $x_4$ shrinks zero at $U=U_{KK}$.
Hence, we are eligible to take T-duality along $x_4$ in the deconfined phase.
We stress that this is not allowed in the original Witten's background with shrinking $x_4$,
and one should take this operation applicable only to the present deconfined phase as
a mere tool for calculating some physical quantities in an easier way.
After T-duality, our $NS5$-brane becomes Type IIB $NS5$-brane because it wraps the $x_4$ direction,
and the $k$ $D2$-brane charges wrapping $x_4$
will transform to $k$ $D1$-brane charges spanning now only $x_1$ and $x_2$. They
are homogeneously distributed in $x_4$ and $S^3\subset S^4$ of our $NS5$-brane world-volume.
We next perform Type IIB S-duality in the system, such that
we finally have a (Euclidean) $D5$-brane wrapping $S^3\subset S^4$ and $x_{4,1,2}$,
with $k$ $F1$-string charges on its world-volume along $x_{1,2}$, homogeneously
distributed in $x_4$ and $S^3\subset S^4$. This can be studied by DBI plus Chern-Simons action.

The T-dualized background of (\ref{decon}) along $x_4$ looks as
\bear
d\tilde s^2&=&\left(U\over R\right)^{3\over 2}\left(f(U)dt_E^2+dx_i^2\right)
+\left(R\over U\right)^{3\over 2}(M_{KK}l_s)^2 d\tilde x_4^2
+\left(R\over U\right)^{3\over 2}\left({dU^2\over f(U)}+U^2 d\Omega_4^2\right)\quad,\nonumber\\
F_5&=&{(2\pi l_s)^3 N\over V_4}d\tilde x_4\wedge \epsilon_4\quad,\quad
e^{\tilde\phi}=e^\phi {1\over\sqrt{g_{44}}}=g_s(M_{KK} l_s)\quad,
\eear
where the parameter $R$ and $f(U)$ are same as before, and more importantly
$\tilde x_4$ has the same period $(2\pi l_s)$ as $x_4$.
One should remember that the relevant rules of T-duality in Ref.\cite{Bergshoeff:1995as}, that is,
$\tilde g_{44}={1\over g_{44}}$ and $e^{\tilde\phi}=e^\phi {1\over\sqrt{g_{44}}}$
must be applied in the coordinate with the fixed period $(2\pi l_s)$\footnote{
Moreover, $C_4^{\rm here}=4 C_4^{\rm there}$ and $C_3^{\rm here}={3\over 2} C_3^{\rm there}$.}.
The resulting 5-form flux simply describes $N$ $D3$-branes homogeneously
distributed along $\tilde x_4$, and the dilation is constant as it should be in a $D3$ background.
After a further S-duality, upon which $ds'^2=e^{-\tilde\phi}ds^2$ and $e^{\phi'}=e^{-\tilde\phi}$,
the final geometry is given by
\bear
ds'^2&=&{1\over g_s(M_{KK} l_s)}\left(U\over R\right)^{3\over 2}\left(f(U)dt_E^2+dx_i^2\right)
+{(M_{KK} l_s)\over g_s}\left(R\over U\right)^{3\over 2} d\tilde x_4^2\nonumber\\
&+&{1\over g_s(M_{KK} l_s)}\left(R\over U\right)^{3\over 2}\left({dU^2\over f(U)}+U^2 d\Omega_4^2\right)\quad,\nonumber\\
F_5&=&{(2\pi l_s)^3 N\over V_4}d\tilde x_4\wedge \epsilon_4\quad,\quad
e^{\phi'} ={1\over g_s(M_{KK} l_s)}\quad,
\eear
and we consider a $D5$-brane wrapping $S^3\subset S^4$ and $x_{4,1,2}$ with $k$ $F1$ charges along $x_{1,2}$.

The (Euclidean) $D5$-brane action is (we omit primes in the above geometry)
\be
S^{D5}_E= \mu_5\int d^6\xi \, e^{-\phi}\sqrt{{\rm det}\left(g^*+ 2\pi l_s^2 F\right)}
-i\mu_5\int \, C_4^{RR}\wedge 2\pi l_s^2 F\quad,
\ee
with $\mu_p=(2\pi)^{-p}l_s^{-(p+1)}$ and we turn on the world-volume gauge flux $F=dA$
only along $x_{1,2}$ to represent $k$ $F1$ charges. To find an expression of $C_4^{RR}$
usable for our purpose
that satisfies $dC_4^{RR}=F_5$ with $F_5$ given above,
we introduce a polar angle $0\le\theta <\pi$ on $S^4$ to write
$d\Omega_4^2=d\theta^2+\sin^2\theta d\Omega_3^2$ and $\epsilon_4=\sin^3\theta d\theta\wedge \epsilon_3$
with $\epsilon_n$ being the volume form of the unit $S^n$. Our $D5$-brane
is wrapping $S^3$ at a constant $\theta$ whose value will be determined dynamically
by a non-zero $k$ $F1$ flux along $x_{1,2}$ on its world-volume.
We choose the gauge such that $C_4^{RR}$ is smooth around $\theta=0$ to have
\be
C_4^{RR}={(2\pi l_s)^3 N\over V_4}
\left(\cos\theta -{1\over 3}\cos^3\theta-{2\over 3}\right) d\tilde x_4\wedge \epsilon_3\quad.
\ee
With these data, it is rather straightforward to compute the above $D5$ action, except a subtle point
that the $k$ $F1$ charge along $x_{1,2}$ is represented by an imaginary $F_{12}\equiv i F$ \cite{Armoni:2008yp}.
One way of seeing this is to consider a Wick-rotated
Lorenzian situation with $x_1$ being the time direction,
where the $F1$ string charge would be unambiguously given by a real $F_{12}$. Upon going back to
our Euclidean situation by the Wick-rotation of $x_1$, the $F1$ flux will transform to
an imaginary value.

The resulting $D5$-brane action density on $x_{1,2}$ after integrating over $S^3$ and $x_4$ is
\be
s_E^{D5}={N\over 4}\sin^3\theta \sqrt{C\left(U\over U_T\right)^3 -F^2}\,+\,{3 N\over 4}
\left(\cos\theta -{1\over 3}\cos^3\theta-{2\over 3}\right)F \quad,
\ee
where $(U,\theta)$ is yet undetermined position of the $D5$-brane, and
\be
C={4^5\pi^6 N^2T^6 \over 3^6 M_{KK}^2 }\quad.
\ee
The $k$ $F1$ charge on its world-volume is now identified as a conserved flux
\be
k= -{\delta s_E^{D5}\over \delta F}= {N\over 4}\left(
{F\sin^3\theta \over \sqrt{C\left(U\over U_T\right)^3 -F^2}}-3\cos\theta+\cos^3\theta +2\right)\quad,
\ee
which can be solved for $F$ in terms of $(U,\theta)$, and the
effective Hamiltonian density one obtains after a Legendre transform with the conserved flux $F$ becomes
\be
h=s_E^{D5}+k F =
{N\over 4}\sqrt{C\left(U\over U_T\right)^3}\sqrt{\sin^6\theta +\left(3\cos\theta-\cos^3\theta
-2+{4k\over N}\right)^2}\quad,
\ee
and we have to minimize this with respect to $(U,\theta)$ to find the $k$-wall tension in our framework.
It is trivial to see that the $D5$-brane sits at the IR end $U=U_T$, while the size of $S^3$ inside
$S^4$ given by the polar angle $\theta$ must be determined by solving the following equation
\be
\sin^2\left(\theta\over 2\right)
 = {k \over N}\quad.
 \ee
As the function of $\theta$ in the left-hand side is a monotonic
function in the range $[0,\pi]$ with values between $(0,1)$, one
checks that $k\leftrightarrow (N-k)$ corresponds to
$\theta\leftrightarrow (\pi-\theta)$ in the solution. With the
solution of $\theta$ and $U=U_T$, the $k$-wall tension is finally
given by \be T_k= {N\over 4}\sqrt{C}\sin^2\theta={8\pi^3\over 27}
N^2\sin^2\theta {T^3\over M_{KK}} ={32\pi^3\over 27} k(N-k)
{T^3\over M_{KK}}\quad, \ee with the desired property of
$T_k=T_{N-k}$. Note that we obtain the precise Casimir scaling
$T_k\sim k(N-k)$, contrary to the $N=4$ SYM result in
Ref.\cite{Armoni:2008yp}. For a small ${k\over N}\ll 1$, one also
confirms that the result reduces to the tension of $k$ $D2$-branes
(\ref{kd2}), \be T_{k}\approx {32\pi^3 \over 27 }kN{T^3\over
M_{KK}}\quad,\quad {k\over N}\ll 1\quad. \ee

\section{Fate of $Z_N$ vacua with fundamental flavors}

In this section, we come to our second motivation, that is, studying
what happens to the $Z_N$-vacua with a presence of fundamental flavors
in a probe approximation, or equivalently in a quenched approximation, via
gauge/gravity correspondence.
We first consider the system of $D3/D7$-branes that is dual to $N=4$ SYM
perturbed by a small number of $N=2$ fundamental hypermultiplets represented by
$N_f$ probe $D7$-branes in a $N$ $D3$-brane background \cite{Karch:2002sh}.
We are especially interested in the dependence on the mass of the flavors given by
the asymptotic distance between $D7$ and $D3$-branes.
We also calculate the energy lift of $Z_N$-vacua in the massless case analytically,
and finally perform a similar analysis in the model by Sakai-Sugimoto \cite{Sakai:2004cn}
for a QCD-like theory with massless chiral quarks.

\subsection{$D3/D7$ system}

Although the appearance of discrete $Z_N$ vacua in the deconefined phase geometry (\ref{n=4decon})
of $N=4$ SYM is a sub-leading effect of large $N$ limit, one can still
advocate a quenched approximation where one neglects back-reaction of
the probe $D7$-branes to the $N=4$ SYM dynamics including the mechanism of
selecting $Z_N$ vacua. We point out that this is just one type of approximation
which is not solely based on the large N limit, because back-reactions of $D7$-branes
may well be of the same sub-leading order of the previous mechanism of $Z_N$ vacua in gravity.
However, we feel that the present quenched approximation
is a useful one to pursue due to its practical calculability and
the possibility of its comparison to lattice QCD in the same quenched approximation.

Working in the quenched approximation in the gravity side means that we take
the previous $Z_N$ vacua of
\be
{1\over 2\pi l_s^2}\int_D\, B^{NS} = {2\pi k\over N}\quad, \label{kvac}
\ee
as a given background and consider the probe dynamics of $D7$-branes embedded in it.
As we neglect possible back-reactions, questions regarding the fate of $Z_N$ vacua
become those of the probe $D7$-branes, such as whether the energy of $k$'th vacuum
is lifted or not. We can address this question by studying the energy of the
probe $D7$-brane in the $k$'th gravity vacuum (\ref{kvac}).

The action of a probe $D7$-brane is
\be
S_E^{D7}=\mu_7\int d^8\xi\,e^{-\phi}\sqrt{{\rm det}\left(g^*+B^{NS*}+2\pi l_s^2 F \right)}\quad,
\ee
where the Chern-Simons term is irrelevant for our purpose.
One should study the equations of motion of the embedding $X^M(\xi)$ and the world-volume
gauge field $F=dA$ in the presence of the background $B^{NS}$ in (\ref{kvac}), and
in general these two are coupled to each other, except the case of trivial vacuum $k=0$
where one can turn off $F=0$ consistently. This special case
was analyzed in Ref.\cite{Babington:2003vm} with varying hypermultiplet mass $m_q$,
where they found a phase transition of meson-melting
at $T_c\sim m_q$. Our problem is a more sophisticated version of theirs.

To parameterize $D7$-brane embedding in the background of (\ref{n=4decon}),
it is convenient to introduce the rectangular coordinate $x^{4,5,6,7,8,9}$
such that $r$ and $S^5$ coordinates are related to them by
\be
dr^2+r^2 d\Omega_5^2 = \sum_{i=4}^9 (dx^i)^2\quad,
\ee
and in fact they are nothing but the original $R^6$ coordinates transverse to the $D3$-branes.
Asymptotically at $r\to\infty$, our $D7$-brane spans four flat directions out of $x^{4-9}$,
say $x^{4-7}$, being a point at $x^{8,9}$ with a distance $2\pi l_s^2 m_q$ from the center.
This fixes the UV boundary condition specified by the mass of the fundamental flavor $m_q$.
As $r$ goes to the infrared region, the $D7$-brane would feel an attraction toward the
horizon at $r=r_H$ and it bends. One easily finds that the spherical symmetry on $x^{4-7}$
is a symmetry of the situation, and the $D7$-brane wraps $S^3$-sphere in $x^{4-7}$ given a
point in $x^{8,9}$.
Moreover $D7$-brane would move only along
one axis on the $x^{8,9}$-plane, and one can take $x^9\equiv 0$ without loss of generality.
Then, the embedding is simply given by a map between $x^8$ and the radius  $\rho$ of $S^3$
inside $x^{4-7}$. One notes that $r^2=(x^8)^2+(x^9)^2+\rho^2$ and
\be
dr^2+r^2 d\Omega_5^2=(dx^8)^2+(dx^9)^2+d\rho^2+\rho^2 d\Omega_3^2\quad,
\ee
so that one can equally describe the embedding by a map between $x^8$ and $r$,
which will be chosen from now on. We will choose the world-volume coordinates $\xi^a$ by
$(t_E,x_i, r, \Omega_3)$, and $x^8$ is a function of $r$ specifying the embedding shape
that is determined dynamically
with the specified UV boundary
condition at $r\to\infty$,
\be
x^8(r\to\infty)= 2\pi l_s^2 m_q\quad,
\ee
as we discuss in the above. The induced metric $g^*_{ab}$ on the $D7$ world-volume is then
\bear
ds_{D7}^2&=&{r^2\over L^2} \left[
\left(1-{\pi^4 T^4 L^8\over r^4}\right)dt_E^2 + dx_i^2\right]+{L^2\over r^2}\left(r^2
-\left(x^8(r)\right)^2\right) d\Omega_3^2\nonumber\\
&+&{L^2\over r^2}
\left[{\left(r-x^8(r){dx^8(r)\over dr}\right)^2\over r^2-\left(x^8(r)\right)^2}
+\left({dx^8(r)\over dr}\right)^2 + {\pi^4 T^4 L^8\over r^4-\pi^4 T^4 L^8}\right]dr^2\quad,
\eear
from which one easily computes the action density along $x_{1,2,3}$ after integrating over $t_E$
and $S^3$,
\be
s_E^{D7}={\beta\over 2^6 \pi^5 g_s l_s^8}\int dr\,\left(r^2-\left(x^8(r)\right)^2\right)^{3\over 2}
\sqrt{A
+\left(B^{NS}_{0 r}+2\pi l_s^2 F_{0 r}\right)^2}\quad,\label{sed7}
\ee
with
\be
A\equiv \left(1-{\pi^4 T^4 L^8\over r^4}\right)
\left[{\left(r-x^8(r){dx^8(r)\over dr}\right)^2\over r^2-\left(x^8(r)\right)^2}
+\left({dx^8(r)\over dr}\right)^2 + {\pi^4 T^4 L^8\over r^4-\pi^4 T^4 L^8}\right]\quad,
\ee
where we turn on $F_{0r}$ along $(t_E,r)$ as a possible world-volume flux
in response to the given background $B^{NS}_{0r}$ with
\be
{\beta \over 2\pi l_s^2}\int_{r_H}^\infty dr \, B^{NS}_{0r}= {2\pi k\over N}\quad.
\ee
We are assuming symmetry along the thermal circle $t_E$ without loss of generality.

The equation of motion of the gauge field is simple to solve; ${\delta s_E^{D7}\over \delta F_{0r}}$
is a constant of motion,
\be
{\delta s_E^{D7}\over \delta F_{0r}}\sim
{\left(r^2-\left(x^8(r)\right)^2\right)^{3\over 2}\left(B^{NS}_{0 r}+2\pi l_s^2 F_{0 r}\right)\over
\sqrt{A
+\left(B^{NS}_{0 r}+2\pi l_s^2 F_{0 r}\right)^2}}\equiv r_H^3 C\quad,\label{conser}
\ee
with some constant $C$ and we factored out $r_H^3$ for later convenience.
One can solve the above for $\left(B^{NS}_{0 r}+2\pi l_s^2 F_{0 r}\right)$,
and the remaining dynamics of the $D7$ embedding $x^8(r)$ will then be described by
the Routhian,
\be
R_E^{D7}=s_E^{D7}-\int dr \,\left({1\over 2\pi l_s^2}
B^{NS}_{0 r}+F_{0 r}\right){\delta s_E^{D7}\over \delta F_{0r}}=
{\beta\over 2^6 \pi^5 g_s l_s^8}\int dr\,\sqrt{A}\sqrt{\left(r^2-\left(x^8(r)\right)^2\right)^3
-r_H^6 C^2}\,,
\ee
given a value of $C$.
The quest is to scan possible range of $C$ and the resulting $D7$-brane embedding shape
from solving the above $R_E^{D7}$, to minimize the original action density $s_E^{D7}$.
Once the final configuration is found, one can compute its energy density from $s_E^{D7}$
to address
the question of energy lift of $Z_N$ vacua.
The necessary numerical analysis, as one varies the quark mass $m_q$,
is beyond the scope of the present work. Identifying a phase transition would be an
interesting future direction.

However, we make several important remarks.
Note that in our coordinate parametrization of $D7$ embedding, there is an inequality
\be
x^8(r) \le r\quad,
\ee
and the equality $x^8(r)=r$ happens precisely when $S^3$ shrinks to zero, where
the $D7$-brane stops going into a lower $r$, that is, it is the end point of the $D7$
embedding in the radial direction $r$.
If this happens at $r=r_0 > r_H$ in the solution, the $D7$-brane doesn't touch the horizon, and
we are in the phase of un-melted mesons in the gauge theory side.
It is important to observe that in this phase
the $D7$-brane covers only part of
the $(t_E,r)$ cigar $D$ of the background (\ref{n=4decon}), that is, a cylinder topology
of $[r_0,\infty]\times t_E$ inside the cigar $D$.
Looking back the expression (\ref{conser}), one readily finds that
the constant of motion $C$ must be zero in this phase considering the point $r=r_0$,
and this in turn says that
\be
{1\over 2\pi l_s^2}
B^{NS}_{0 r}+F_{0 r} \equiv 0\quad,\label{triv}
\ee
in all region of the $D7$-brane world-volume.
Then the analysis of the embedding shape
and the resulting energy density from $s_E^{D7}$ becomes precisely
equal to the case of trivial vacuum $k=0$ or without $B^{NS}$/$F$ fluxes at all.
One concludes that $Z_N$-vacua are not lifted and persist to exist in this phase
in quenched approximation.
One interprets that the world-volume flux $F$ nullifies the background $B^{NS}$ flux
to minimize $s_E^{D7}$. In the gauge theory point of view,
as the world-volume gauge symmetry on the $D7$-brane is the flavor global symmetry
in the gauge theory side, this points out an interesting phenomenon of
a dynamically generated monodromy of flavor symmetry along the thermal circle that
counteracts the existing $Z_N$ Wilson line in the $Z_N$-vacua, to make the $Z_N$-vacua
viable even in the presence of fundamental flavors.
One mathematical remark is that (\ref{triv}) is possible precisely
because the topology of $D7$-brane in $(t_E,r)$ is cylindrical, so that
there is no topological restriction of its $\int F$ value.

In the other phase where $x^8(r) < r$ remains true until it hits the horizon $r=r_H$,
the $D7$-brane meets the horizon and the mesons are melted.
A crucial point in this phase is that the $D7$-brane now wraps the whole cigar $D$ of $(t_E,r)$
with topology of a two-plane, and there is an important topological restriction
\be
\int_D \, F = 2\pi m\quad,\quad m\in Z\quad.\label{rest}
\ee
Suppose one tries to put $C=0$, or equivalently ${1\over 2\pi l_s^2}
B^{NS}_{0 r}+F_{0 r} \equiv 0$, to minimize the action $s_E^{D7}$ as before, then
\be
\int_D\, F=-{1\over 2\pi l_s^2}\int_D\,
B^{NS}=-{2\pi k\over N}\, \notin \,2\pi Z\quad,
\ee
except the trivial vacuum $k=0$. In other words, in the $k$'th vacuum
it is not possible to completely nullify the $Z_N$ Wilson line $B^{NS}$
due to a topological restriction, and one has a non-vanishing $C\ne 0$ or
${1\over 2\pi l_s^2}
B^{NS}_{0 r}+F_{0 r} \ne 0$ in the non-trivial $Z_N$ vacua.
For each integer $m$, $C$ must be determined to satisfy (\ref{rest}), and one should
choose $m$ that minimizes the energy $s_E^{D7}$.
Looking at the expression (\ref{sed7}) of $s_E^{D7}$, one sees
that the resulting energy is always higher than the trivial vacuum $k=0$ where
${1\over 2\pi l_s^2}
B^{NS}_{0 r}+F_{0 r} \equiv 0$, and we conclude that
the $Z_N$ vacua are lifted in the phase of melted mesons.

The question at which phase the system finds its solution given the
UV boundary condition of the flavor mass $m_q$ should be addressed numerically,
but for a sufficiently large ${m_q\over T}\gg 1$ it is natural to expect
the former phase where the mesons are un-melted and the $Z_N$ vacua survive,
while in the extreme case of $m_q=0$ it is obvious that $x^8(r)\equiv 0$ and
we are in the latter phase of melted mesons with $Z_N$-vacua lifted.
One naturally expect a phase transition to happen at some point $T_c\sim m_q$.
An interesting question is whether the critical temperature $T_c$ depends on $k$ or not
because of the additional contribution from the $B^{NS}$/$F$ fluxes to $s_E^{D7}$,
whose answer is beyond the present paper.

In the simplest case of $m_q=0$ where one can consistently set $x^8(r)\equiv 0$
to be in the melted-meson phase, it is possible to  compute the
energy lift of the $Z_N$-vacua analytically.
Note that we have $A=1$ in (\ref{sed7}) with $x^8(r)\equiv 0$, and
the equation of motion for the gauge field (\ref{conser}) simplifies as
\be
{r^3 \left(B^{NS}_{0 r}+2\pi l_s^2 F_{0 r}\right)\over
\sqrt{1
+\left(B^{NS}_{0 r}+2\pi l_s^2 F_{0 r}\right)^2}}\equiv r_H^3 C\quad,
\ee
which can be solved for $F$,
\be
F_{0r}=-{1\over 2\pi l_s^2}B_{0r}+{C\over 2\pi l_s^2\sqrt{\left(r\over r_H\right)^6-C^2}}\quad.\label{sol}
\ee
As discussed before, $C$ can not take any continuous values, but is restricted to only
discrete values to satisfy the topological constraint (\ref{rest}).
Explicitly, integrating both sides of (\ref{sol}) over $D$, one has
\be
2\pi m = -{2\pi k\over N}+{C\beta\over 2\pi l_s^2}\int_{r_H}^\infty dr\, {1\over
\sqrt{\left(r\over r_H\right)^6-C^2}}
=-{2\pi k\over N}+{C\beta r_H \over 2\pi l_s^2}\int_{1}^\infty d\hat r\, {1\over
\sqrt{\hat r^6-C^2}}\quad,
\ee
so that given $(m,k)$, the $C$ is determined by
\be
C\int_1^\infty
d\hat r\, {1\over
\sqrt{\hat r^6-C^2}}={C\over 2}\,{_2F_1}\left({1\over 3}, {1\over 2},{4\over 3},C^2\right)=
2\pi\left(m+{k\over N}\right){1\over \sqrt{\pi g_s N}}\quad,\label{ceq}
\ee
where we have used $r_H=\pi T L^2=\pi T \left(4\pi g_s N\right)^{1\over 2}l_s^2$.
Once $C$ is determined as a function of $(m,k)$, it is straightforward to
calculate the energy density $s_E^{D7}$,
\be
s_E^{D7}-s_E^{D7}\Big|_{k=0}
= {\beta r_H^4\over 2^6 \pi^5 g_s l_s^8}\int_1^\infty d\hat r
\, \left({\hat r^6\over \sqrt{\hat r^6 -C^2}}-\hat r^3\right)
={\pi g_s N^2 T^3\over 4}\int_1^\infty d\hat r
\, \left({\hat r^6\over \sqrt{\hat r^6 -C^2}}-\hat r^3\right)\quad,\label{finen}
\ee
where we have subtracted the value of the trivial vacuum $k=0$ to see the energy lift of
the $(m,k)$ state. For a fixed $k$, one should choose $m$ that minimizes the energy
to finally find the energy lift of the $k$-vacuum,
\be
\Delta \epsilon_k={\rm min}_{m\in Z}\left(s_E^{D7}\Big|_{(m,k)}-s_E^{D7}\Big|_{k=0}\right)\quad.
\ee
From (\ref{finen}), one has to minimize $C^2$, and an inspection of (\ref{ceq})
shows that this is achieved by the smallest value of $|m+{k\over N}|$.
Therefore, our result is invariant under $k\to k+N$ and $k\to (N-k)$,
which are the desired properties due to $Z_N$ nature and charge conjugation.
For low lying $k$-vacua with ${k\over N}\ll 1$, we have $m=0$ to minimize the energy,
and an easy calculation gives us
\be
\Delta \epsilon_k \approx {\pi^2}{k^2\over N} T^3\quad,\quad {k\over N}\ll 1\quad.\label{resultd3d7}
\ee
One also finds that $\Delta\epsilon_k$ has a discontinuous slope at ${k\over N}={1\over 2}$.

\subsection{The model of Sakai and Sugimoto}

The Sakai-Sugimoto model is obtained by introducing probe $D8/\bar{D8}$ branes in the Witten's
$D4$-brane background, that span $S^4\times R^{3,1}\times U$ and are point-like
in $x_4$-direction. In the original weak-coupling
$D4/D8/\bar{D8}$-brane picture, one finds left(right)-handed chiral quarks from $D4-D8$($D4-\bar{D8}$)
string spectrum, and at low energy the resulting 4D gauge theory flows precisely
to QCD
with massless chiral flavor quarks. One therefore expects the gravity picture to capture some of the
interesting dynamics of this semi-realistic gauge theory, at least for low enough energy regime.
In the confined geometry (\ref{confinedgeo})
where $(x_4,U\ge U_{KK})$ makes a cigar shape
with vanishing $x_4$ at $U=U_{KK}$, the $D8/\bar{D8}$-brane pair must join with each other at IR,
realizing spontaneous chiral symmetry breaking in an intuitively geometric way.
Our current interest is however on the opposite case of the deconfined geometry (\ref{decon}),
where $x_4$ remains finite everywhere, and we instead have a cigar shape $D$ with $(t_E,U\ge U_T)$.
In this case, each $D8$ and $\bar{D8}$-brane separately
meets the horizon at $U=U_T$ and their world-volume embedding
is simply parameterized by a constant $x_4$ position. We will consider a single pair of
$D8/\bar{D8}$-branes for simplicity.

As we consider $D8/\bar{D8}$-branes in the $Z_N$-vacua of the background,
\be
{1\over 2\pi l_s^2} \int_D B^{NS}= {\beta\over 2\pi l_s^2} \int_{U_T}^\infty dU\,
B^{NS}_{0U}={2\pi k\over N}\quad,
\ee
one can easily find that the embedding shape can still  be
consistently given by a constant $x_4$ from the equation of motion, and we will take this
without further detail. One can also be convinced by translational symmetry along $x_4$.
Because the Chern-Simons term of the $D8/\bar{D8}$ doesn't play any role here,
the actions of $D8$ and $\bar{D8}$ are identical, and the total action is two times
the DBI action of the $D8$-brane,
\be
S_E=2\mu_8\int d^9\xi\, e^{-\phi}\sqrt{{\rm det}\left(g^*+B^{NS*}+2\pi l_s^2 F\right)}\quad,
\ee
where we choose the world-volume coordinates $\xi^a$ to be $(t_E,x_i,U,\Omega_4)$.
It is straightforward to compute the above, turning on a possible world-volume gauge field $F_{0U}$
along the cigar $D$ of $(t_E,U)$. After integrating over $S^4\times t_E$, the action
density in the spatial $R^3$-direction becomes
\be
s_E = {N^{1\over 2}\over 3\cdot 2^4 \pi^{11\over 2}g_s^{1\over 2} l_s^{15\over 2}
T} \int_{U_T}^\infty dU\,U^{5\over 2}\sqrt{1+\left(B^{NS}_{0U}+2\pi l_s^2 F_{0U}\right)^2}\quad.
\ee
The equation of motion for $F_{0U}$ is simply ${\delta s_E\over \delta F_{0U}}={\rm const.}$,
which we parameterize by
\be
{U^{5\over 2}\left(B^{NS}_{0U}+2\pi l_s^2 F_{0U}\right)\over \sqrt{1+
\left(B^{NS}_{0U}+2\pi l_s^2 F_{0U}\right)^2}}\equiv U_T^{5\over 2} C\quad,
\ee
with a dimensionless constant $C$, and from this one easily solves for $F_{0U}$ as
\be
F_{0U}=-{1\over 2\pi l_s^2} B_{0U} +{1\over 2\pi l_s^2}{C\over \sqrt{\left(U\over U_T\right)^5-C^2}}\quad.
\ee

As in the previous case of $D3/D7$ system, an important point is a topological restriction
\be
\int_D\, F = \beta\int_{U_T}^\infty dU\, F_{0U}=2\pi m\quad,\quad m\in Z\quad,
\ee
and this determines the constant $C$ in terms of given $(k,m)$ as follows,
\be
C\int_{1}^\infty d\hat U {1\over \sqrt{\hat U^5 -C^2}}={2C\over 3}\,{_2F_1}\left({3\over 10},{1\over 2}
,{13\over 10},C^2\right)=
\left(m+{k\over N}\right){4\pi^2 l_s^2\over \beta U_T}
=\left(m+{k\over N}\right){9\over 4\pi g_s N l_s T}\,,
\ee
where ${_2F_1}$ is the hypergeometric function.
Once $C$ is given, one can easily calculate the difference in the action density $s_E$
from the $k=0$ vacuum where $\left(B^{NS}_{0U}+2\pi l_s^2 F_{0U}\right)=0$, and after
sorting out the coefficients, the final answer is
\be
s_E-s_E\Big|_{k=0}={2^{10} \pi^5 g_s^3 N^4 l_s^3 T^6\over 3}
\int_1^\infty d\hat U\, \hat U^{5\over 2}\left(\sqrt{\hat U^5\over \hat U^5-C^2}-1\right)\quad.
\ee
For a given $k$-vacuum, one should find the integer $m$ such that it minimizes
the above result, and that is the final energy lift of the $k$'th vacuum $\Delta\epsilon_k$.
Because $C^2$ is a monotonic increasing function of $\Big|m+{k\over N}\Big|$, one should
take the smallest possible value of $\Big|m+{k\over N}\Big|$, and this again gives us
$\Delta \epsilon_k=\Delta \epsilon_{k+N}$ and $\Delta \epsilon_{k}=\Delta \epsilon_{N-k}$.
The slope is discontinuous in the middle ${k\over N}={1\over 2}$.
For small ${k\over N}\ll 1$, one has $m=0$ and one can easily find that
\be
C\approx {27\over 8\pi}{1\over g_s N l_s T} {k\over N} \quad,\quad {k\over N}\ll 1\quad,
\ee
and
\be
\Delta \epsilon_k\approx 6^4\pi^3 g_s l_s k^2 T^4 =2^3 3^4 \pi^2 \left(g_{YM}^2 N\right)
{k^2\over N}{T^4\over M_{KK}}\quad,\quad {k\over N}\ll 1\quad,
\ee
where we have used $g_s l_s={g_{YM}^2\over 2\pi M_{KK}}$.
This is qualitatively different from the case of $D3/D7$ system (\ref{resultd3d7}).

\section{Conclusion}

In this work, we first study $k$-walls with $k\sim N$ in the Witten's $D4$-brane
background of pure $SU(N)$ Yang-Mills theory in large N limit, largely
motivated by a recent work in $AdS_5\times S^5$ by Armoni, Kumar and Ridgway \cite{Armoni:2008yp}.
We propose and check consistency of the picture that $k$ (Euclidean) $D2$-branes
blow up into a $NS5$-brane wrapping $S^3$ inside $S^4$ of the background via Myers effect.
We compute the tension of $k$-wall by performing suitable U-duality to transform the $NS5$-brane
into a $D5$-brane. Our result is a precise Casimir scaling behavior of the $k$-wall
tension $T_k\sim k(N-k)$. We note that
the same Casimir scaling was also obtained for the $k$-quark flux tube tension
in Ref.\cite{Callan:1999zf}.

Our second subject is to consider fate of $Z_N$-vacua in the presence of
fundamental flavors in quenched approximation via gauge/gravity correspondence.
In our study of $D3/D7$ system, we point out
an interesting phenomenon of phase transition as one varies $m_q\over T$,
which separates the phase of un-melted mesons with $Z_N$-vacua survived and the phase
of melted mesons with $Z_N$-vacua lifted.
An interesting question is whether the critical temperature depends on $k$,
which should be addressed by a numerical analysis in the future.

In the special case of $m_q=0$,
we calculate the energy lift of $k$'th vacua analytically.
Our result is consistent with the $Z_N$-nature and the charge conjugation where $k\to (N-k)$.
We also find a tower of stable states parameterized by integers $m$
above the lowest energy state.
We do the similar analysis in the Sakai-Sugimoto model for an approximate QCD-like theory.

\vskip 1cm \centerline{\large \bf Acknowledgement} \vskip 0.5cm

We would like to thank Adi Armoni, S. Prem Kumar and Jefferson M. Ridgway for
helpful discussions, and Andrei Smilga for clarifying comments on correct Euclidean interpretation of
the $Z_N$-walls.
We also acknowledge Alberto Guijosa, Mohammad Edalati and Rene Meyer
for helpful comments.

 \vfil

\end{document}